# Shaped two-photon excitation deep inside scattering tissue

Eirini Papagiakoumou[1], Aurélien Bègue[1], Osip Schwartz[2], Dan Oron[2] and Valentina Emiliani[1]

[1]*Neurophysiology and New Microscopies Laboratory, Wavefront engineering microscopy group, CNRS UMR 8154, INSERM U603, Paris Descartes University, 45 Rue des Saints Pères, 75270 Paris Cedex 06, France*

[2]*Department of physics of complex systems, Weizmann institute of science, Rehovot 76100, Israel*

*Corresponding author*: Valentina Emiliani**,** Neurophysiology and New Microscopies Laboratory, Wavefront engineering microscopy group, CNRS UMR 8154, INSERM U603, Paris Descartes University, 45 Rue des Saints Pères, 75270 Paris Cedex 06, France, tel: +33142864253, e-mail: valentina.emiliani@parisdescartes.fr

*Classification*: BIOLOGICAL SCIENCES, Neuroscience

**Abstract**

Light is the tool of the 21$^{st}$ century. New photosensitive tools offer the possibility to monitor and control neuronal activity from the sub-cellular to the integrative level. This ongoing revolution has motivated the development of new optical methods for light stimulation. Among them, it has been recently demonstrated that a promising approach is based on the use of wavefront shaping to generate optically confined extended excitation patterns. This was achieved by combining the technique of temporal focusing with different approaches for lateral light shaping including low numerical aperture Gaussian beams, holographic beams and beams created with the generalized phase contrast method. What is needed now is a precise characterization of the effect of scattering on these different methods in order to extend their use for in depth excitation.

Here we present a theoretical and experimental study on the effect of scattering on the propagation of wavefront shaped beams. Results from fixed and acute cortical slices show that temporally focused spatial patterns are extremely robust against the effects of scattering and this permits their three-dimensional confinement for depths up to 550 μm.

\body

**Introduction**

The combination of light microscopy with functional reporters (1-4), caged compounds and, more recently, optogenetics (5-7) offers the possibility to control activation and inhibition of neuronal activity and monitor functional responses in a non-invasive manner enabling the analysis of well-defined neuronal population within intact neuronal circuits and systems. Interestingly, those tools have permitted to address key biological questions with relatively simple illumination methods using widefield visible light illumination (1, 8-11). However, some limitations in the specificity of genetic targeting (different cellular types can express the same optogenetic channels) and the intricate morphology of the brain (sub-cellular compartments, such as dendrites and axons, can reach regions far away from the cell soma), make it challenging to, for example, individuate subsets of genetically identical interconnected cells, or to establish the role of specific spatiotemporal excitatory patterns in guiding animal behavior. To reach such degree of specificity, more sophisticated illumination methods are required, permitting control of light patterning deep inside tissue. In this respect the use of visible light illumination is problematic due to the lack of optical sectioning and, for scattering samples, the low penetration depth.

Two-photon (2P) excitation mitigates the effects of scattering due to the longer excitation wavelength and permits superior axial resolution due to the nonlinear dependence of the absorption probability on the excitation intensity. However, a major drawback in the use of 2P photoactivation is that the number of activatable molecules/channels within the small excitation volume of a standard 2P diffraction-limited spot (1-3 $\mu m^3$) might be insufficient e.g. to generate an action potential (12-14) or a detectable fluorescence signal from genetic reporters.

Several solutions have been recently proposed to this issue. One is based on increasing the excitable areas using low numerical aperture (NA) beams focused on a single spot (12) or scanned through multiple locations (13). However, the rapid loss of axial and, for scanning, temporal resolution limits their use to small excitation areas. To increase the excitation area without affecting the axial resolution, alternative solutions including static beam multiplexing (15) or low-NA temporally focused (16) Gaussian beams have been demonstrated (17). However these approaches are limited to fixed excitation patterns. Increased flexibility to generate, for example, excitation patterns adapted to the shape of a sub-cellular compartment, or to cover multiple cell somata can be reached by quick scanning of a temporally focused beam (17, 18). However, in this case a compromise has to be found between axial resolution and scanning time.

Alternative to these approaches full parallel methods for scanless multiscale excitation have been demonstrated. The first one, originally proposed for optical tweezers (19) and then for patterned 1P uncaging (20-22), consists in using the principle of digital holography (DH). With this approach, dynamical reconfiguration of the optical wavefront allows efficient multiple-scale excitation going from single to multiple diffraction limited spots, to shapes covering a single subcellular process or a large population of sparse neurons. In 2P excitation this approach combined with TF allowed to reach an axial resolution of few μm independently of the excitation pattern (23-25).

An alternative approach for lateral light patterning is based on the technique of the generalized phase contrast (GPC) (26). The main advantage of GPC is to avoid spatial intensity fluctuations, which are intrinsic to DH (27). With this approach combined with TF, 2P depth-resolved speckle-free arbitrary patterns can be generated and efficient 2P excitation of ChR2 in cultured neurons and brain slices could be demonstrated (28).

Overall these results show that enhanced 2P photoactivation of optogenetic actuators and reporters can be reached by using extended axially confined excitation patterns (29). These can be generated by using two complementary approaches based on serial or parallel excitation. Both methodologies have limits and advantages and each of them might prove to be best suited for different experimental configuration. For parallel approaches, a fundamental question that still needs to be answered, in order to be fully characterized, is how wave-shaped beams (low-NA Gaussian beams, holographic and GPC generated temporally focused beams) propagate deep into scattering media.

Recently, Dana et al. (30) have compared the propagation in scattering tissue of a low-NA Gaussian beam with TF to that of tightly focused Gaussian beams. Using both Monte Carlo simulations and experiments they found that axial sectioning provided by TF deteriorates more rapidly with propagation depth than in 2P line-scanning microscopy. Yet, it permits a reasonably good optical sectioning also for low-NA beams.

Perhaps even more important than the effect of scattering on the axial resolution is its effect on the predetermined shape of the excitation pattern. Generally, scattering turns a homogeneously illuminated region into a speckled pattern. Yet, when photoexciting through turbid tissues, it is of utmost importance to maintain the spatial shape of the excitation beam.

In this paper, we explore the propagation through fixed and acute brain slices of low-NA Gaussian beams, holographic and GPC generated temporally focused beams. We show that TF excitation does not only maintain a reasonable degree of axial confinement through scattering tissue, but is also relatively robust against scattering, maintaining the general spatial excitation profile irrespective of the particular realization of scattering inside the sample. This enables accurate and controlled photoexcitation of predetermined volumes up to hundreds of microns deep into scattering specimens.

**Results**

**Theoretical modeling**

In order to elucidate the evolution of temporally focused beams inside turbid tissue, we start by considering the results of numerical simulations of light propagation within a scattering phantom. In the simulations, we used 100 fs pulses centered at either 800 nm or 950 nm, as in the experiments performed on biological preparations. As a scattering sample, we consider a homogeneous medium in which dielectric spheres of 2 μm in diameter, having a refractive index higher by 0.1 than the surrounding medium and an average concentration of 1 scatterer per 1000 μm$^3$, are randomly distributed.

In Figure 1, we compare the 800 nm 2P excitation pattern generated at the focal plane of the focusing lens after propagation through 200 μm of the turbid sample (corresponding to approximately 1.5 effective scattering lengths, as defined by the integrated signal decay exponent (31)) for three different techniques of generation of the excitation pattern: a low-NA Gaussian beam (Fig. 1a) with a full width at half-maximum (FWHM) of 7 μm, a 10 μm circular spot generated with GPC (Fig. 1b) and a 10 μm circular spot generated with DH (Fig. 1c). For comparison, the left column shows the excitation pattern without scattering. The middle column shows the excitation pattern with scattering but without TF, while in the right column we plot the excitation pattern with TF. Evidently, standard illumination (no TF) results in significant speckle, hardly reminiscent of the original shape for either the Gaussian excitation pattern or the GPC-generated spot. The situation is slightly better in the case of the holographic beam, since we already start with a speckled intensity distribution with this technique. In contrast, temporally focused excitation yields relatively smooth 2P excitation patterns, particularly along the x-axis (along which, colors are dispersed). This observation is, again, more evident in the case of originally homogeneous excitation patterns (Gaussian, Fig. 1a, and GPC beams Fig. 1b), as compared with holographic illumination (Fig. 1c), where the effect is partially suppressed by the existence of inherent speckles.

The simulations thus suggest that TF significantly reduces the distortion of multiphoton excitation patterns when illuminating through scattering tissue. In the following, we set out to show this experimentally for photoexcitation through brain slice samples, and also consider the effect of scattering on the degree of axial confinement of photoexcitation in these samples.

### Propagation of low-NA Gaussian beams

To evaluate the scattering effect experimentally, we first tested the propagation of low-NA Gaussian beams through brain slices of different thickness by using a "double microscope" configuration (Fig. 2a and Methods). We excited fluorescence at 800 nm on a thin fluorescent layer through fixed slices of 50-250 μm thickness. In Figure 2b, we show the fluorescence images obtained by using a Gaussian beam of 11 μm FWHM when the beam is directly imaged on the fluorescent layer (left panel) and when it propagates through a brain slice of 250 μm thickness, without (middle panel) and with TF (right panel). Clearly, the experimental results resemble the numerical simulation predictions (Fig. 1a): in the absence of TF, the excitation pattern ends up with large speckles whose position varied with the location on the slice, whereas the shape for temporally focused beams remains much better preserved. The predicted additional smoothing for TF beams along the x-axis is also visible.

To evaluate the effect of scattering on the axial resolution we performed the axial scanning (see Methods) of the TF Gaussian beam through different thicknesses of scattering tissue and we found, in agreement with the theoretical predictions and the data reported by Dana et al. (30), that the axial resolution started to deviate from its value without scattering (2.7±0.3 µm) for slice thicknesses of the order of the measured scattering length, $\ell_s$= 170±12 µm.

**Propagation of GPC beams**

To characterize the effects of scattering in the case of a GPC beam, we repeated the same experiments as for low-NA Gaussian beams by using a round spot of 15 µm generated with a GPC optical setup. In Figure 3a, we show how the spot is imaged when it propagates through brain tissue. Significant distortion is seen already after 50 µm of propagation without TF and the circular shape is barely visible after 250 µm. In contrast, the temporally focused GPC beam retained, with incredible precision, the original shape, even through 250 µm of tissue. From the comparison between Figure 2b and Figure 3a we can also see that GPC temporally focused beams retain their form better than temporally focused Gaussian beams. To understand this, one must consider the fact that distortions arise primarily from the interference of weaker scattered waves with the stronger unscattered ones, which serve as a "local oscillator" to amplify their effect. The Gaussian illumination pattern is relatively smooth, so that this amplification occurs throughout the entire pattern. In contrast, in the GPC excitation spot, having relatively sharp edges, light scattered outside the shape is too weak to induce 2P absorption.

In Figure 3b, the experimental (orange dots) values for the axial resolution as a function of the slice thicknesses are shown, and are in good agreement with the simulation (black dots) prediction. The axial resolution starts to deviate from the one measured without scattering (2.8±0.1 µm) for depths near the scattering length, which, for the series of samples used in this case, was $\ell_s$= 135±12 µm (Fig. S1a). Scattering for smaller depths, resulted only in an increased signal at the tails of the axial propagation curves (Fig. S1b).

The ability of GPC beams to retain with a high precision the original excitation shape was also tested for two patterns containing fine details (Fig. 3c). A first pattern was generated based on the fluorescence image of a layer V cortical neuron (28) (top) and a second one on the base of the fluorescence image of a dendritic branch of a Purkinje cell (bottom). These results show that temporally focused GPC beams preserved their original form even for fine structures with μm-precision, in stark contrast with the results obtained for GPC beams without TF.

**Propagation of holographic beams**

Finally, we measured the propagation of holographic generated beams, by repeating the same experiments using a round spot of 15 μm generated with a DH optical setup. We found that, up to a slice thickness of 250 μm, holographic beams are quite robust against scattering even without TF (Fig. 4a,b). Yet, a high axial resolution is only obtained for TF-DH (23), exhibiting more or less a similar behavior as in the TF-GPC case.

It should be noted, however, that unlike TF-GPC, fine structures generated with DH quickly deteriorate upon propagation in the scattering medium (Fig. S2a). What is more interesting in the case of DH is the ability to switch the excitation configuration from diffraction-limited spots to multiple large spots with the advantage in the latter case to retain the original shape unaffected by scattering (Fig. S2b).

Similar to the results obtained for TF-GPC beams, we found that the axial resolution for temporally focused holographic beams remains almost unchanged with respect to the value measured without scattering (3.7±0.3 μm) for thicknesses shorter than the scattering length, $\ell_s$, and increases of about a factor of 5 for thicknesses of the order of twice the $\ell_s$.

As previously reported (20, 23, 32), holographic beams, even without TF, show improved axial resolution than low-NA Gaussian beams thanks to their particular phase distribution at the back aperture of the objective. We evaluated how this axial behavior is affected by scattering in that case, and found that axial resolution is almost unaffected up to brain thickness of 250 μm (Fig. S3).

**Propagation through acute slices**

In order to characterize the maximum depth at which we can use patterned photostimulation in live tissues, we tested the propagation of GPC and holographic patterns through acute cortical brain slices (250-550 μm thick) prepared from juvenile rats. As in the experiments with fixed slices, we studied the effects of scattering on the excitation shape and axial resolution by recording the images on a thin fluorescent layer after propagation through brain slices of different thickness. For these experiments, we used an excitation wavelength at 950 nm, which not only represents the optimal excitation wavelength for ChR2 (13, 17, 28), but should also mitigate the effects of scattering due to the longer excitation wavelength, as can be seen in the simulations plotted in Figures 3b and 4b.

In Figure 5a, we show the images of the GPC excitation pattern shown in Figure 3c after propagation through acute slices of 300 and 550 μm. Although, as expected the overall intensity decreases, the fine structure of the pattern is still preserved after 550 μm of tissue with remarkable accuracy for the temporally focused beam. In contrast, the original shape is completely lost in absence of TF (Fig. 5a, right panel).

In Figure 5b, the same experiment is performed for the propagation of a 15 μm holographic spot. In agreement with the results obtained for fixed slices, DH beams appear extremely robust to scattering. TF-DH transmitted shapes exhibit almost no distortion even after 550 μm, while for DH without TF, the original shape starts to be affected at around 300 μm.

The average FWHM values measured for the GPC and holographic patterns are shown in Figure 5c. For the GPC beam, the axial resolution has been measured on the part of the excitation shape mimicking the soma or the principal dendrite. In the latter case, we found, as expected, values slightly better than the ones measured on the soma, since fine structures resemble the case where the grating is illuminated with a line, giving an axial resolution closer to the point-scanning limit (24).

Overall these results clearly show the benefits of TF in maintaining axial resolution for shaped excitation patterns in scattering media.

**DISCUSSION**

We have investigated the propagation, through fixed and acute brain slices, of extended patterns generated by three different methods: low-NA Gaussian beams, beams generated with the GPC method and holographic beams. For the three configurations, we compared the propagation with and without TF.

We also presented a theoretical model based on Fresnel propagation in a scattering phantom that reproduces well the experimental transmitted shapes and axial resolution for each of these configurations.

Overall, we have found that opposite to standard illumination, TF permits to fully maintain the original lateral shape at least up to 500 μm. Moreover, the original shape is maintained irrespective of the particular details of the scattering medium. In order to understand this remarkable difference, one should consider the effect of scattering in the case of multicolor photoexcitation. For standard illumination (no TF), the speckle pattern of all the colors comprising the excitation pulse are nearly identical (due to the fact that the spectral bandwidth is much smaller than the carrier wavelength, $\Delta\lambda/\lambda=0.01 \ll 1$) (33). Thus, the excitation pattern practically reflects a single realization of a speckle pattern generated by the medium.

For temporally focused excitation the situation is significantly different. Since different excitation colors traverse different paths through the scattering medium, their respective speckle patterns are nearly uncorrelated, providing significant smoothing to the overall 2P response. Moreover, speckle patterns of 'nearby' colors, which partially traverse the same path but at a different propagation direction, are partially correlated but spatially shifted along the x-axis. This effect provides additional smoothing along the x-axis, which is apparent in the right column of Figures 1a and 1b. The latter effect (termed 'Smoothing by Spectral Dispersion' or SSD) has long been used for smoothing out speckle in inertial confinement fusion applications (34). The combined effect leads to overall illumination patterns which strongly resemble the original intended shape. Overall, these effects lend a self-healing quality to the temporally focused excitation profile.

In the case of temporally focused digital holograms, where the excitation pattern is inherently speckled (23, 24), the overall shape is also maintained upon the use of TF but the excitation pattern is not smoothed out (see Fig. 1c).

By measuring the axial propagation for the three different methods we found, in agreement with simulation predictions, that the axial resolution is well maintained for thickness of the order of the scattering length, $\ell_s$. For greater depths, the axial resolution starts to deteriorate reaching a value approximately 5 times larger at $2\ell_s$.

The comparison between the three different approaches has revealed that among these, a low-NA Gaussian beam results in a somewhat worse axial resolution. The degree of deterioration is in good agreement with results presented in the study of Dana et al. for similar experimental conditions (30).

TF-GPC beams permit to conserve impressively well fine details up to depths of 550 μm in acute slices. However in GPC constraints on the excitation density practically limit the maximum achievable excitation field (28). TF-GPC beams therefore represent the most suitable choice for excitation in small fields of view and of fine subcellular processes. Moreover the absence of speckles allows extending the use of this technique for imaging.

For TF-DH beams, the unavoidable presence of speckles makes it difficult to conserve fine structures through brain thicknesses greater than 50 μm. However for uniform shapes (single or multiple large spots) the original shape remains basically unaffected after propagation through slices of up to 550 μm. DH beams without TF are also very robust to scattering conserving the original shape up to 300 μm. Interestingly, the axial resolution seems to deteriorate somewhat less sharply than for TF-GPC beams and, although slightly worse, is well conserved also in the absence of TF. With DH, contrary to GPC, large excitation fields and efficient multi-scale excitation (from single diffraction-limited spots to multiple-cell excitation) are possible using the same optical setup, which makes this configuration the best choice for applications requiring this flexibility.

Two existing alternatives to TF for photoexcitation through scattering media are Bessel beams and the use of adaptive optics. 'Bessel beams' (35) reconstruct their original shape regardless of the particular realization of scattering inside the medium and have recently been used for a variety of applications, including particle trapping (36) and light-sheet microscopy (37). However a major disadvantage of these non-diffracting beams, in the context of photostimulation, is the complete absence of optical sectioning (due to the lack of diffraction) and that the excitation shape is limited to a circular geometry.

Adaptive optics provides another alternative for improved propagation through scattering samples. Indeed, it has recently been shown that photoexcitation of a diffraction-limited spot (38-40) or even a set of spots deep inside scattering media, is possible by use of adaptive wavefront correction (41). These means are, nevertheless, both slow and sensitive, as a different correction has to be applied at any given spatial position within the sample, and little data has been reported on their efficacy in correcting the propagation of extended shaped patterns (42).

From the above considerations, we find that the use of DH or GPC with TF in combination with amplified ultrafast laser pulses of at least a few microjoules (available from either continuously pumped amplifiers or cavity dumped oscillators) is probably the method of choice for extending 2P optogenetics to *in vivo* applications.

**Methods**

**Numerical simulations**

For the numerical simulation of the evolution of temporally focused beams inside turbid tissue we used 100 fs pulses centered at either 800 nm or 950 nm. As a scattering sample we considered a homogeneous medium in which 2 μm dielectric spheres, having a refractive index higher by 0.1 than the surrounding medium and an average concentration of 1 scatterer per 1000 μm$^3$, are randomly distributed. The simulations are performed iteratively, whereby in each simulation step the beam is propagated via the angular spectrum of plane waves approach and then a spatial phase is imprinted on it according to the distribution of scattering spheres. This approach, where backscattering is neglected, is a legitimate approximation within, for example, brain slices, where the index contrast within the sample is typically very small.

**Optical setup for beam propagation characterization**

The light source used for the experiments was a mode-locked Ti:Sapphire laser (Mai Tai, Spectra-Physics; Δt=100 fs, Δλ=10 nm, repetition rate 82 MHz), whose output beam power was controlled by an achromatic half-wave (λ/2) plate combined with a Glan-laser calcite polarizer cube.

The characterization of the beam propagation through the different types of brain slices was done in a microscope configuration with two opposite placed objectives ("double microscope") (Fig. 2a), described in previous works (20, 23). 2P fluorescence was excited through a 60x microscope objective (Olympus, LUMPLFL60xW/IR2, NA 0.90) on a ~1 μm-thick fluorescent layer of rhodamine-6G in PMMA, spin-coated on a glass coverslip, through different thicknesses of fixed or acute brain slices. The emitted fluorescence was collected by a second objective (Olympus, UPLSAPO60xW, NA 1.20) placed opposite to the excitation one, disposing a correction collar for the coverslip thickness. The use of two independent objectives allowed straightforward imaging of out-of-focus planes of the excitation volume, in order to precisely derive the axial propagation of the excitation beam around the objective's focal plane. The imaging objective was fixed and focused on the thin fluorescent layer, while the upper one was moved along the axial direction (±40 μm around its focal plane, in steps of 0.5 μm) with a piezo-scanning stage working in closed-loop (MIPOS100SG, Piezosystem Jena). The fluorescence was imaged by a collection tube lens (f=150 mm) to a CCD camera (CoolSNAP HQ2, Roper Scientific). To reject the excitation light, an emission filter (Chroma Technology HQ 535/50M) and a dichroic filter (Chroma Technology 640DCSPXR) were placed in front of the CCD camera.

To deduce the axial resolution, we scanned the excitation objective through the fluorescent layer and measured the excited fluorescence intensity at each plane. The axial resolution was considered as the FWHM value of the plot of the normalized 2P fluorescence intensity for the different axial planes.

*Temporal focusing setup*

The accordingly modulated laser beam for the generation each time of the Gaussian, holographic or GPC patterns was sent to a reflectance diffraction grating (830 lines/mm) aligned perpendicular to the optical axis of the microscope. The illumination angle of the grating was ~42º at 800 nm, or 52° at 950 nm, such as the +1 order diffracted beam at the center frequency of the excitation pulse was directed along the optical axis of the microscope. The grating dispersed the various frequency components of the beam, which were imaged onto the sample via a telescope comprised of an achromatic lens (f=500mm) and the excitation objective. Whenever experiments without temporal focusing were performed, the diffraction grating of the optical setup was replaced with a plane metallic mirror.

*Gaussian beam illumination*

The laser beam was spatially expanded with a variable beam expander (2x-5x) and sent to the diffraction grating for temporal focusing. The expansion of the laser beam before the grating was such as the lateral size of the Gaussian beam at the sample plane was 11 μm at FWHM.

*Digital Holography-generated beam illumination*

The principle of DH consists in calculating with an iterative Fourier transform algorithm (43, 44), given a target intensity at the objective's focal plane, the phase pattern at the rear aperture of the objective that permits one to reproduce the target. The calculated phase-hologram is addressed to a spatial light modulator (SLM) that imposes the phase modulation onto the input beam's wavefront. After propagation through the objective, the beam is focused onto a spot, reproducing straight away the desired template.

In the experimental setup the laser beam was expanded (10x) in order to match the input window of a Liquid Crystal on Silicon Spatial Light Modulator (LCOS-SLM) (Hamamatsu Photonics X10468-02), illuminated at oblique incidence, as in previous works (23, 24). As described there, the beam's wavefront at the SLM plane was modified by using a custom-designed software (20) that, given a target intensity distribution at the focal plane of the microscope objective, calculates the corresponding phase-hologram by using the iterative algorithm. The beam reflected from the LCOS-SLM was Fourier transformed by a 1000 mm focal length achromatic lens and formed a first image of the target intensity onto the grating for TF. In this way a depth-resolved temporally focused excitation pattern was generated at the focal plane of the objective, whose lateral spatial distribution was a reduced (~1/170) replica of the pattern generated on the grating.

*Generalized Phase Contrast-generated beam illumination*

In GPC the output light distribution is obtained by the interference between a signal and an unscattered light component (zero order), which travel along the same optical axis. The desired target intensity map is converted into a spatially similar phase map that is addressed on the SLM. A phase contrast filter (PCF) placed at the Fourier plane of the SLM imposes a half-wave ($\lambda/2$) phase retardation between the on-axis focused component and the higher-order diffracted Fourier components. The interference between the phase-shifted focused and the scattered light, allows generating a pure phase-to-intensity conversion at the output plane.

Here, the LCOS-SLM was controlled by another option of the custom-designed software that, given a target intensity distribution at the focal plane of the microscope objective converts the intensity map into a binary phase map and addresses the output profile to the SLM. The beam reflected from the SLM was separated in its Fourier components by a 400 mm focal length achromatic lens and focused on the PCF, positioned at the Fourier plane of this lens. The on-axis, low spatial frequency components were shifted in phase by $\lambda/2$, by the PCF and then, the second 300 mm achromatic lens recombined the high (signal wave) and low (synthetic reference wave) spatial frequency components. The introduced phase shift caused these components to interfere and produced an intensity distribution according to the spatial phase information carried by the higher spatial frequencies. Then the beam was sent to the diffraction grating and the setup for temporal focusing was following. The 500 mm lens and the microscope objective formed the second 4-f lens system that scaled the intensity distribution (~ 1/110) on the sample plane.

For further details on how to get optimal contrast conditions at the output plane of the system (focal plane of the 300 mm lens, which coincides with the grating plane) and how to choose the size of the PCF, the reader can refer to our previous work (28).

**Brain slices preparation**

All experiments followed European Union and institutional guidelines for the care and use of laboratory animals (council directive 86/609EEC). See SI Appendix for preparation of fixed and acute slices.

**Data acquisition and analysis**

Optical sectioning and image acquisition have been performed using the Metamorph software (version 7.0, Molecular Devices). Acquisition time was 200-300 ms for all fluorescence images. Data analysis was performed with Image J (version 1.43u), Origin 8.0 (OriginLab) and Matlab R2009a. Experimental data are reported as mean ± standard deviation (SD).

**Acknowledgments**

The authors would like to thank J. Montanaro, G. Bouchery and A. Schorscher-Petcu for the preparation of fixed slices and M. Briand for her contribution to the data analysis. EP was supported by the "Fondation pour la Recherche Médicale" (FRM). AB was supported by a doctoral fellowship from Paris School of Neuroscience (ENP). VE was supported by the Human Frontier Science Program (RGP0013/2010) and the "Fondation pour la Recherche Médicale" (FRM équipe). OS and DO acknowledge support by the European Research Council starting investigator grant SINSLIM and the Crown center of photonics.

**Figure Legends**

**Fig. 1.** Numerical Modeling. Multiphoton excitation pattern generated at the focal plane of the objective by numerical simulation of the propagation of three different excitation beams, (**a**) a low-NA Gaussian beam, (**b**) a GPC-generated round spot and (**c**) a spot generated with DH, through 200 μm of a turbid medium consisting of 2 μm dielectric spheres randomly distributed, having a refractive index higher by 0.1 than the surrounding medium and an average concentration of 1 per 1000 μm$^3$. The images of the beams not scattered (left column), scattered but not temporally focused (middle column) and scattered when they are temporally focused (right column) are illustrated. Simulations were performed at 800 nm, with 100 fs laser pulses. Scale bars: 10 μm.

**Fig. 2.** Propagation of low-NA Gaussian beams. (**a**) Schematic layout of the "double microscope" experimental configuration used for the characterization of beam propagation through different thicknesses of brain tissue. (**b**) 2P excited fluorescence

images at the focal plane of the excitation objective of the "double microscope" of a low-NA Gaussian beam (11 μm FWHM) without scattering (left) and after propagation through 250 μm of fixed cortical brain slices, without the use of TF (middle) and when TF was used (right). λ=800 nm; maximum laser power used at the sample plane: 14 mW (without TF) – 140 mW (with TF); scale bars: 10 μm.

**Fig. 3.** Propagation of GPC beams. (**a**) 2P excited fluorescence images at the focal plane of the excitation objective of a 15 μm GPC-generated spot without scattering (left panel) and after propagation through 50 and 250 μm of fixed cortical brain slices, without TF in the optical setup (middle panel) and when TF was used (right panel). λ=800 nm. (**b**) Variation of the FWHM of the axial intensity distribution of the GPC spot in respect to the scattering depth. Black and blue dots correspond to simulations at 800 and 950 nm respectively, whereas orange dots correspond to experimental data (average values for at least 4 measurements in different locations). (**c**) 2P excited fluorescence images at the focal plane of the excitation objective of GPC-generated excitation patterns mimicking a neuron with its small processes (top) and the dendrites of a Purkinje cell (bottom) without scattering (left column) and after propagation through 250 μm of fixed coronal cortical rat brain slices. In the middle, images without TF in the optical setup are shown, while on the right, images are taken when TF was used. Laser power used at the sample plane: 40-145 mW. The gray-scale level is normalized to the peak intensity of the fluorescence image without scattering. Scale bars: 10 μm.

**Fig. 4.** Propagation of holographic beams. (**a**) 2P excited fluorescence images at the focal plane of the excitation objective of a 15 μm holographic spot without scattering (left image) and after propagation through 50 and 250 μm of fixed cortical brain slices, without TF in the optical setup (top) and when TF was used (bottom) (λ=800 nm). (**b**) Variation of the FWHM of the axial intensity distribution of the holographic spot in **a**, in respect to the scattering depth. Black and blue dots correspond to theoretical modeling at 800 and 950 nm respectively, whereas orange dots correspond to experimental data. Laser power used at the sample plane: 14-420 mW. The gray-scale level is normalized to the peak intensity of the fluorescence image without scattering. Scale bars: 10 μm.

**Fig. 5.** Propagation through acute slices. (**a**) 2P excited fluorescence images at the focal plane of the excitation objective of the GPC-pattern mimicking the neuron with its small processes. From left to right: without scattering, after propagation through acute cortical brain slices of 300 μm with TF, 550 μm with TF and 550 μm without TF in the optical setup. (**b**) 2P excited fluorescence images at the focal plane of the excitation objective of a 15 μm holographic spot. From left to right: without scattering, after propagation through acute cortical brain slices of 300 μm with TF, 550 μm with TF and 550 μm without TF in the optical setup. (**c**) Variation of the FWHM of the axial intensity distribution in respect to the scattering depth. Red and green dots correspond to the FWHM values measured on the soma and the apical dendrite respectively, of the excitation shape shown in **a**, orange dots correspond to FWHM values measured for the spot in **b**. $\lambda$=950 nm; laser power used at the sample plane: 60-230 mW. The gray-scale level is normalized to the peak intensity of the fluorescence image without scattering. Scale bars: 10 μm.

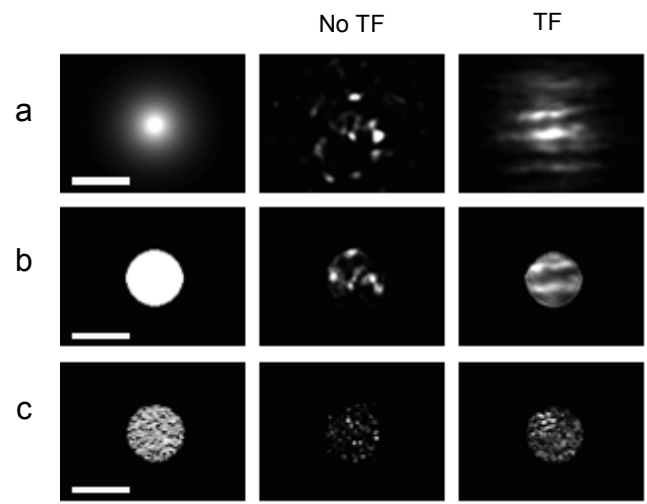

a

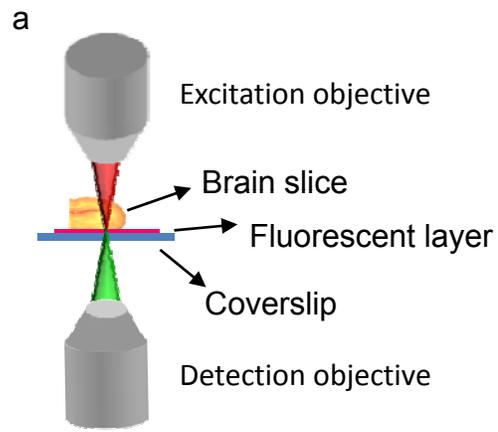

b

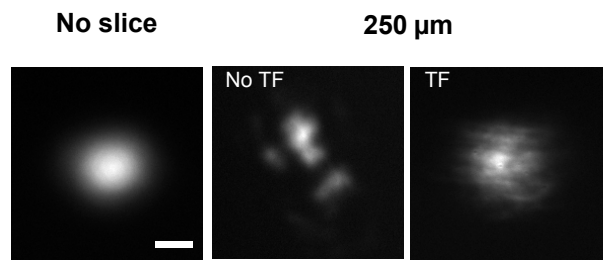

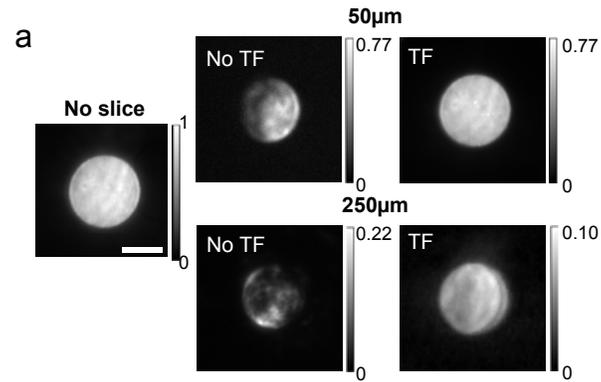

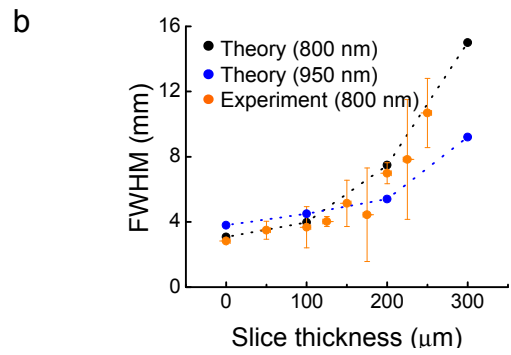

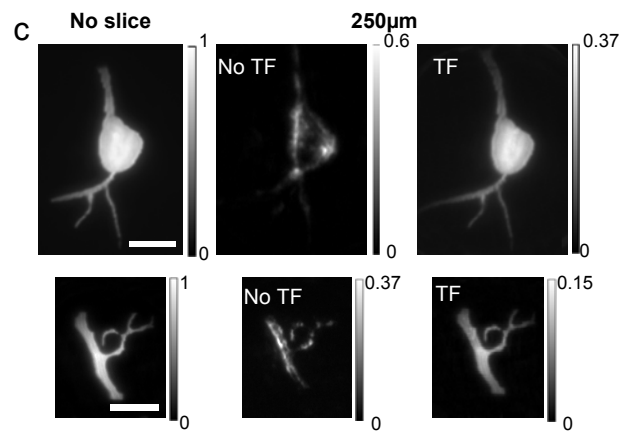

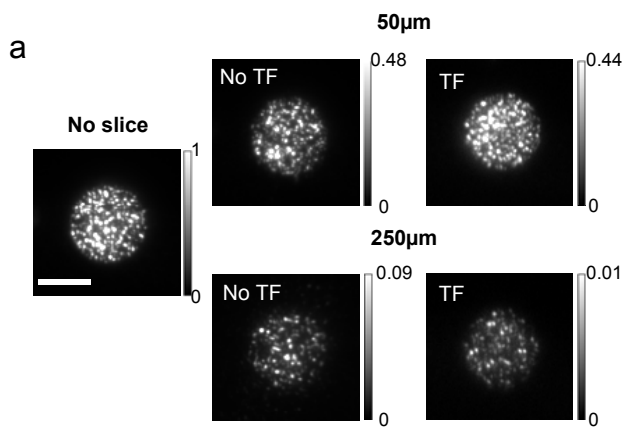
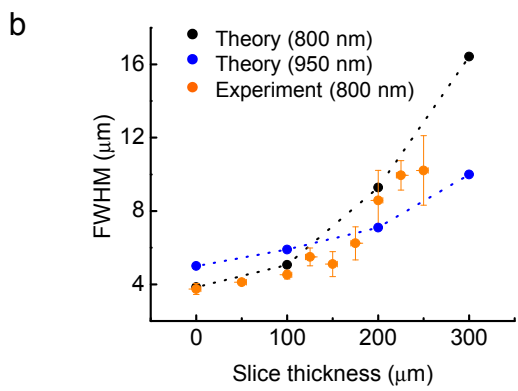

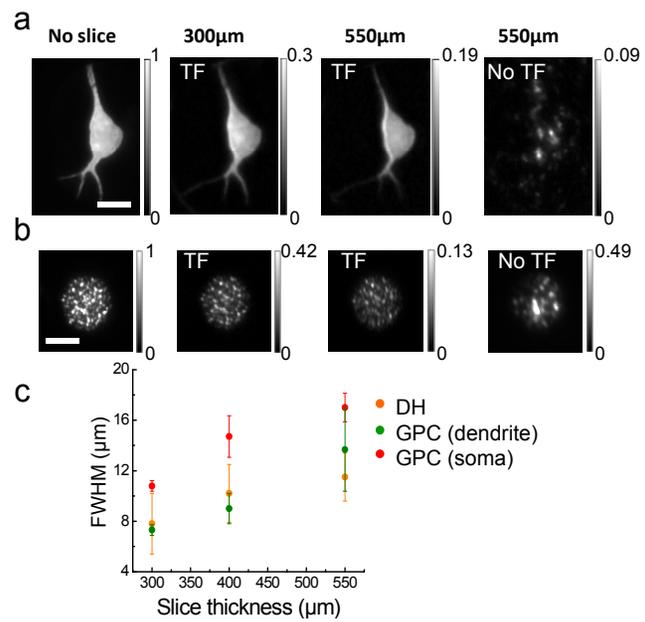